%% file: arxiv_paper.tex
\documentclass{article} 
\usepackage{iclr2026_conference,times}

\input{math_commands.tex}

\usepackage{graphicx}
\usepackage{hyperref}
\usepackage{booktabs}
\usepackage{amsmath}
\usepackage{amssymb}
\usepackage{url}
\usepackage{multirow}
\usepackage{cleveref}
\usepackage{multirow}
\usepackage[table]{xcolor}
\usepackage{subcaption}

\title{UniFlow-Audio: Unified Flow Matching for Audio Generation from Omni-Modalities}

\iclrfinalcopy

\author{Xuenan Xu$^{1,\dag,}$\thanks{Equal contribution. $^\dag$ Project Lead.}~~, Jiahao Mei$^{2,5,*}$, Zihao Zheng$^{1,2}$, Ye Tao$^{1,2}$, Zeyu Xie$^3$, Yaoyun Zhang$^2$, \\ \textbf{Haohe Liu$^4$, Yuning Wu$^5$, Ming Yan$^5$, Wen Wu$^1$, Chao Zhang$^1$, Mengyue Wu$^2$} \\
$^1$Shanghai Artificial Intelligence Lab, $^2$Shanghai Jiao Tong University\\
$^3$Peking University, $^4$Meta, $^5$Alibaba Group\\
\texttt{wsntxxn@gmail.com} \\
}

\begin{document}

\maketitle
    
\begin{abstract}
Audio generation, including speech, music and sound effects, has advanced rapidly in recent years.
These tasks can be divided into two categories: time-aligned (TA) tasks, where each input unit corresponds to a specific segment of the output audio (e.g., phonemes aligned with frames in speech synthesis); and non-time-aligned (NTA) tasks, where such alignment is not available.
Since modeling paradigms for the two types are typically different, research on different audio generation tasks has traditionally followed separate trajectories.
However, audio is not inherently divided into such categories, making a unified model a natural and necessary goal for general audio generation.
Previous unified audio generation works have adopted autoregressive architectures, while unified non-autoregressive approaches remain largely unexplored.
In this work, we propose UniFlow-Audio, a universal audio generation framework based on flow matching.
We propose a dual-fusion mechanism that temporally aligns audio latents with TA features and integrates NTA features via cross-attention in each model block.
Task-balanced data sampling is employed to maintain strong performance across both TA and NTA tasks.
UniFlow-Audio supports omni-modalities, including text, audio, and video.
By leveraging the advantage of multi-task learning and the generative modeling capabilities of flow matching, UniFlow-Audio achieves strong results across 7 tasks using fewer than 8K hours of public training data and under 1B trainable parameters.
Even the small variant with only $\sim$200M trainable parameters shows competitive performance, highlighting UniFlow-Audio as a potential non-auto-regressive foundation model for audio generation.
Code and models will be available at \url{https://wsntxxn.github.io/uniflow\_audio}.
\end{abstract}

\section{Introduction}
\label{sec:intro}

With the rapid evolution of generative models~\citep{vaswani2017attention,ho2020denoising,lipman2023flow}, recent works have achieved remarkable improvements in generation quality~\citep{esser2024scaling,polyak2024movie}, promoting popularity of artificial intelligence generated content (AIGC).
As an important modality, audio has also made remarkable progress in various generation tasks, with text-to-speech synthesis (TTS)~\citep{wang2023neural} and text-to-audio (T2A) generation~\citep{liu2023audioldm} serving as representative tasks.
Traditional audio generation models are designed for specific tasks, such as converting text to speech or music. 
This paradigm is suboptimal, as it overlooks the interconnected nature of real-world auditory information.

\begin{figure}[ht]
    \centering
    \includegraphics[width=0.95\linewidth]{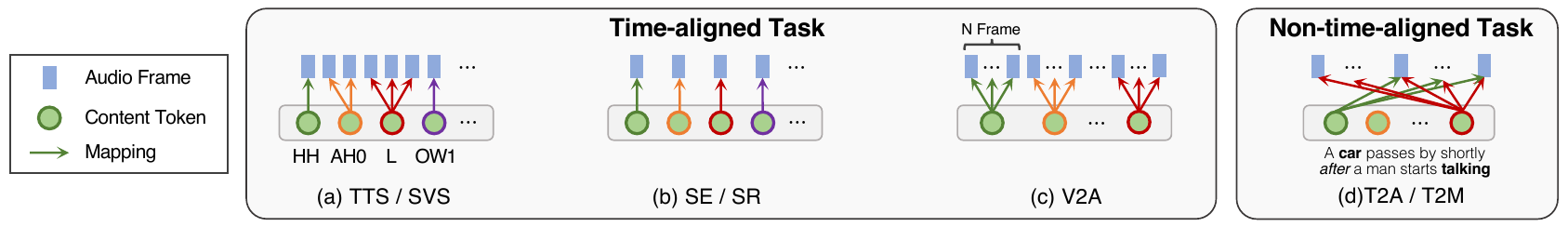}
    \caption{Illustration of time-aligned (TA) tasks and non-time-aligned (NTA) tasks.}
    \label{fig:TA_NTA}
\end{figure}

To overcome this limitation, we aim at a unified framework for audio generation that accommodates diverse input (text, audio, video) and output modalities (speech, music, sound effect).
We observe that despite their differences, these tasks can be fundamentally categorized by the temporal relationship between input and output: either \textit{\textbf{time-aligned (TA)}} or \textit{\textbf{non-time-aligned (NTA)}}, as shown in \Cref{fig:TA_NTA}.
For TA tasks, there is strict temporal alignment between input and output, such as the monotonic alignment in text-to-speech (TTS), the one-to-one frame alignment in speech enhancement (SE), and one-to-$N$ frame alignment in video-to-audio~(V2A).
In contrast, NTA tasks, such as T2A, do not require such a temporal alignment constraint: the input sequence (textual description) corresponds holistically to the entire output soundscape, with semantic consistency being the primary objective rather than temporal correspondence.
This fundamental difference in alignment requirements has historically necessitated specialized modeling approaches for TA and NTA tasks.


While recent works have explored unified audio generation with autoregressive (AR) architectures, unified non-autoregressive (NAR) approaches remain relatively underexplored.
UniAudio~\citep{yang2024uniaudio} adopts an AR paradigm, achieving strong zero-shot performance on both AR and NAR tasks.
However, AR models rely on sequential decoding and discrete tokenizers, whereas NAR models generate continuous audio representations in parallel, which may offer advantages in latency and quality.
Thus, NAR-based unified audio generation remains worth exploring.
AudioX represents an NAR attempt, but it focuses exclusively on NTA tasks and cannot handle TA tasks such as TTS, which require variable-length generation.
Meanwhile, task-specific NAR models like VoiceFlow~\citep{guo2024voiceflow} perform well on TA tasks by temporally aligning content embeddings with audio latents, yet this modeling paradigm does not generalize to NTA tasks.
This leaves a gap for a single NAR framework capable of unifying both TA and NTA tasks within one modeling paradigm.

In this work, we propose UniFlow-Audio, a universal audio generation framework based on flow matching that unifies both TA and NTA tasks within a single non-auto-regressive (NAR) model.
From the modeling perspective, we propose a dual-fusion mechanism to temporally align audio latents with input features for TA tasks, while utilizing cross-attention to integrate input features for NTA tasks, ensuring high-quality generation across both categories.
To avoid interference between the two fusion strategies, task-irrelevant features (\textit{i.e.}, NTA features for TA tasks and TA features for NTA tasks) are replaced with learnable dummy embeddings, keeping TA and NTA feature integration disentangled.
Both TA and NTA tasks are integrated in each block of the backbone (block-wise fusion), enabling the input to more effectively guide the generation.
To balance the amount of different data types, we adopt a task-balanced sampling strategy to balance the ratio between TA and NTA data during training. 
Moreover, UniFlow-Audio supports a broader range of input modalities than prior works, including text, audio, and visual signals.
With all these modalities and tasks involved, UniFlow-Audio learns the shared knowledge across different tasks, which in turn yields competitive or superior performance compared to task-specific baselines.
Notably, compared with other unified audio generation models (see \Cref{subsec:unified_audio_generation} for details on data and model sizes), our small variant ($\sim$200M parameters), trained on fewer than 8K hours of public data, achieves strong results, underscoring the data efficiency and parameter effectiveness achieved by UniFlow-Audio.

The contributions of this work can be summarized as follows:
\begin{enumerate}
    \item We provide a novel perspective that formulates diverse audio generation tasks through temporal alignment. 
    \item We propose UniFlow-Audio, the first flow-matching-based universal audio generation framework that unifies TA and NTA tasks.
    \item We design model architectures and data sampling strategies to balance TA and NTA tasks while ensuring the generation quality, including a dual-fusion mechanism, block-wise fusion, and task-balanced sampling.
    \item UniFlow-Audio achieves strong results with limited open-source data and parameters on a variety of tasks, demonstrating the advantages of a unified audio generation model. 
    \item We open-source the code and model to provide a potential unified NAR audio generation foundation model, enabling further theoretical exploration and practical applications.  
\end{enumerate}

\section{Related Work}

\paragraph{Unified Audio Generation}
\label{subsec:unified_audio_generation}
Recently, the research paradigm in audio generation has shifted from task-specific models to unified frameworks capable of handling multiple tasks within a single model.
Such frameworks facilitate cross-domain knowledge sharing and improve data efficiency.
Representative works include UniAudio~\citep{yang2024uniaudio} and AudioX~\citep{tian2025audiox}.
UniAudio is a large language model (LLM) based AR model that discretizes audio and various input modalities into token sequences and leverages a multi-scale Transformer to model inter- and intra-frame correlations.
UniAudio is trained on 165K hours of data.
Despite 11 tasks being included, the video input modality is not supported in UniAudio.
In contrast, AudioX adopts an NAR Diffusion Transformer (DiT) with a multi-modal input masking strategy to enhance robustness and generation performance.
While trained on 29K hours of large-scale curated data, it focuses exclusively on NTA tasks.
Compared with these pioneering works, UniFlow-Audio proposed a flow-matching-based unified NAR framework that achieves good performance on both TA and NTA tasks, with omni input modalities involved (text, audio, video) whilst trained on smaller datasets.

\paragraph{Flow Matching for Audio Generation}
Recent NAR generative models, diffusion models~\citep{ho2020denoising} and flow matching~\citep{lipman2023flow}, have attracted significant attention in audio generation due to their strong generative capabilities and the fast inference speed through parallel generation.
NaturalSpeech2~\citep{shen2024naturalspeech}, E3-TTS~\citep{gao2023e3}, and AudioLDM~\citep{liu2023audioldm} demonstrate the capabilities of latent diffusion models on speech and audio generation.
To achieve high-fidelity generation with extremely few steps, flow matching is adopted for T2A and TTS with low latency~\citep{eskimez2024e2,chen2025f5,guan2024lafma}.
It alleviates the high inference latency inherent to the iterative denoising process in diffusion models by directly learning a continuous velocity field that transports noise into data in a few integration steps, rather than requiring a substantial number of discrete denoising iterations.
Flow matching is also employed in hybrid TTS systems such as CosyVoice~\citep{du2024cosyvoice} to refine acoustic details given discrete tokens predicted by the AR component.
Motivated by the success of flow matching in prior speech and audio generation works, UniFlow-Audio adopts flow matching as the backbone.

\section{UniFlow-Audio}

\begin{figure*}
    \centering
    \includegraphics[width=\linewidth]{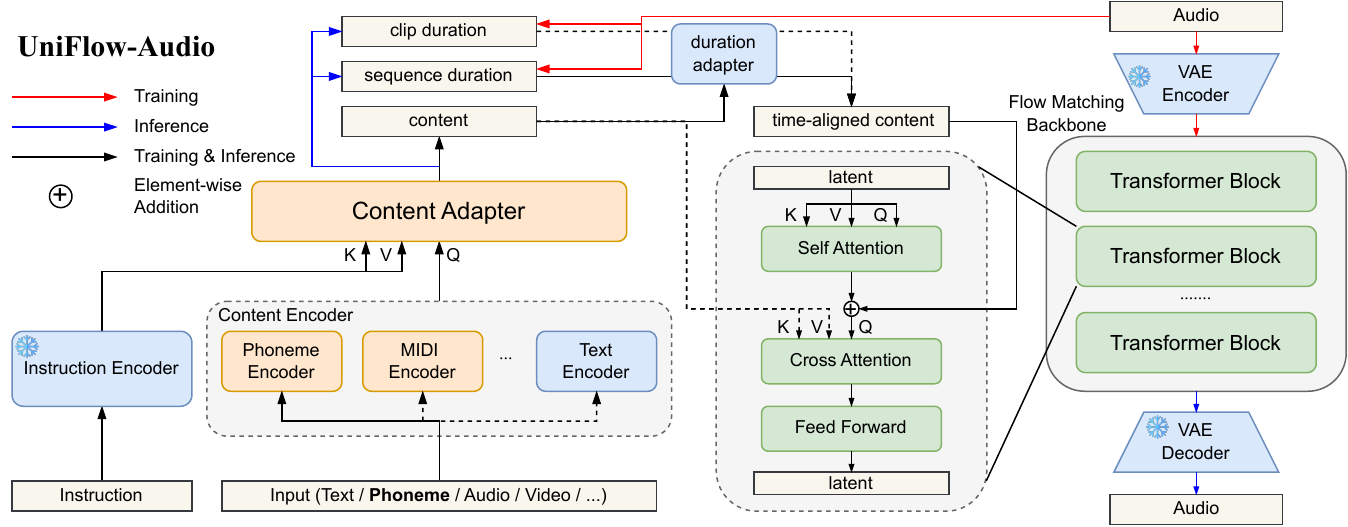}
    \caption{Overview of UniFlow-Audio. The content encoder and adapter transform the input and task instruction into content embedding. Based on the predicted duration, the content embedding is expanded to time-aligned content embedding. A dual-fusion mechanism is applied: the latent is fused with the content by cross attention, and fused with time-aligned content by addition.}
    \label{fig:framework}    
\end{figure*}

As \Cref{fig:framework} shows, UniFlow-Audio is a unified flow-matching-based audio generation framework that consists of four parts: a variational autoencoder (VAE) that compresses the raw long audio signal into a short sequence, a content encoding part for extracting features from the input content and task instruction, a duration adapter that generates TA content embeddings, and a Transformer-based flow matching backbone.

\subsection{Audio Representation for Generation}
Following \citep{evans2025stable}, we employ a VAE that operates on raw waveforms for direct waveform generation and reducing latency.
The VAE encoder compresses the waveform $\mathbf{x} \in \mathbb{R}^{L}$ into a latent representation $\mathbf{A} \in \mathbb{R}^{L/2^R\times D}$, where $L$, $R$ and $D$ denote the waveform length, compression ratio and latent dimension, respectively.
The VAE architecture also follows \citep{evans2025stable}, with details shown in \Cref{subsec:vae_details}.
We train the VAE on a mixture of high-quality speech, music, singing voice and general audio datasets to improve the generation performance on various domains.


\subsection{Content Encoding with Task Instruction}
All inputs are transformed into continuous embeddings $\mathbf{C}$ instead of discrete tokens to avoid information loss by modality-specific content encoders:

\textbf{Phoneme \& MIDI}: 
For TTS, phonemes from grapheme-to-phoneme conversion (g2p)\footnote{\url{https://github.com/MontrealCorpusTools/Montreal-Forced-Aligner}} and x-vectors~\citep{wang2023wespeaker} for speaker information are used as input.
We use the Transformer-based encoder from FastSpeech2~\citep{ren2020fastspeech} as the content encoder.
Singing voice synthesis (SVS) is similar to TTS, except that the input is MIDI rather than phonemes.
In addition to phoneme embeddings, the MIDI encoder incorporates pitch, pitch duration, and slur information, which are fused with the phoneme embeddings through addition.

\textbf{Text}: For T2A and text-to-music generation (T2M), the input is a coarse text description without the alignment information. 
We use Flan-T5~\citep{chung2024scaling} as the encoder following~\citep{majumder2024tango,evans2025stable}.

\textbf{Audio}: For audio input, we reuse the VAE as the encoder to compress the sequence length.

\textbf{Video}: For video input in video-to-audio generation (V2A), we use CLIP~\citep{radford2021learning} combined as the encoder.

The VAE, Flan-T5, and CLIP are frozen during training.
After obtaining $\mathbf{C}$ from the content encoder, we further integrate task instructions to inject explicit task-specific information, enabling the model to distinguish between tasks that share the same input modality (\textit{e.g.}, T2A and T2M).
This integration is achieved through an instruction encoder and a content adapter: the former maps the textual instruction into embeddings $\mathbf{I}$, and the latter fuses $\mathbf{C}$ with $\mathbf{I}$ via cross-attention (Attn) by
\begin{equation}
    \mathbf{C^I} = \text{Attn}(\mathbf{C}, \mathbf{I}, \mathbf{I}) + \mathbf{C}.
\end{equation}
Regarding each task, we design 10 diverse textual instructions that describe the objective (details shown in \Cref{sec:task_instructions}).
During training, one instruction is randomly selected from each task as the input, whereas during inference, a fixed instruction is used.

With task-involved content embeddings $\mathbf{C}^{\mathbf{I}}$, a clip duration $d_c \in \mathbb{R}^{+}$ and a sequence duration $d_s \in (\mathbb{R}^{+})^{L}$ are predicted.
Since UniFlow-Audio is an NAR model, both TA and NTA tasks rely on $d_c$ to determine the output length.
$d_s$ is only required by TA tasks for duration adaptation, which will be introduced in \Cref{subsec:duration_adaptation}.
For the duration predictor, we adopt the architecture in FastSpeech2. 

\subsection{Duration Adapter}
\label{subsec:duration_adaptation}
As introduced in \Cref{sec:intro}, audio generation tasks can be divided into TA and NTA categories by their temporal alignment constraint.
In NTA tasks where input and target audio lack temporal correspondence, cross-attention mechanism is typically used to integrate $\mathbf{C}$ into the generation process.
In TA tasks, alignment information is often explicitly leveraged for generation.
For instance, TTS relies on phoneme-to-frame alignment to expand linguistic units, while speech enhancement (SE) inherently operates on frame-aligned noisy and clean audio pairs.
In such cases, content embeddings are aligned and concatenated with audio features, a process that may require a duration adapter. 

Building on this insight, we introduce a unified \textit{duration adapter} to explicitly align content embeddings with audio latents across all TA tasks.
We posit that this explicit alignment offers superior efficacy for TA tasks than the implicit mechanisms of cross-attention.
Specifically, $\mathbf{C^I}$ is expanded to a time-aligned content $\mathbf{C^I_T}$. That is,
\begin{equation}
\mathbf{C^I_T}  = \big[\ \underbrace{c^I_1, \ldots, c^I_1}_{(d_s)_1},\ \underbrace{c^I_2, \ldots, c^I_2}_{(d_s)_2}, \ \dots, \ \underbrace{c^I_N, \ldots, c^I_N}_{(d_s)_N} \ \big].
\end{equation}
Based on the sequence duration $d_s$, the duration adapter repeats each embedding $c^I_i$ in $\mathbf{C^I}$ for $(d_s)_i$ steps, producing $\mathbf{C^I_T}$ that matches the length of the audio latents.
For TTS and SVS, $d_s$ specifies the number of audio latents per phoneme.
For SE and V2A, each value in $d_s$ is fixed, since each input audio latent or video frame corresponds to a fixed number of target audio latents.
For NTA tasks, $d_s$ is set to a constant dummy value to achieve a unified design.
During training, ground-truth durations are used to obtain $\mathbf{C^I_T}$.

\subsection{Dual-Fusion Flow Matching Transformer}
\label{subsec:dual_fusion}
The generation backbone is a flow-matching Transformer composed of multiple Transformer blocks.
We employ a dual-fusion mechanism to integrate both $\mathbf{C^I}$ and $\mathbf{C^I_T}$ into generation.
Within each block, the audio latent $\mathbf{A}$ is first processed by self-attention.
The flow matching timestep $\tau$ is incorporated by adaptive layer norm (AdaLN) as
\begin{equation}
\mathbf{A} = (\text{AdaLN}_{\text{SA}}\circ\text{Attn})(\mathbf{A}, \mathbf{A}, \mathbf{A}).
\end{equation}
Next, $\mathbf{C^I_T}$ is fused with $\mathbf{A}$ by addition, as they are temporally aligned. That is,
\begin{equation}
    \mathbf{A} = \mathbf{A} + \mathbf{C^I_T}.
\end{equation}
Finally, as in standard Transformer blocks, cross-attention with $\mathbf{C^I}$ and feed-forward network (FFN) are applied to obtain the block output by
\begin{align}
    \mathbf{A} &= \text{Attn}(\mathbf{A}, \mathbf{C^I}, \mathbf{C^I}) + \mathbf{A} \\
    \mathbf{A} &= (\text{AdaLN}_{\text{FFN}}\circ\text{FFN})(\mathbf{A}).
\end{align}

To prevent interference between the two fusion streams, we replace the ineffective input with learnable dummy embeddings.
Specifically, $\mathbf{C^I}$ for TA tasks and $\mathbf{C^I_T}$ for NTA tasks are set as dummy embeddings.

\subsection{Training and Inference}

We train the model using the flow matching loss, which encourages the velocity field $v_\theta(\mathbf{z}_\tau, \tau)$ to match a target velocity field, so that the continuous-time flow induced by $v_\theta$ transports data latents $\mathbf{z}_0 \sim p_\text{data}$ to a standard Gaussian $\mathbf{z}_1 \sim \mathcal{N}(0, \mathbf{I})$:
\begin{align}
\frac{d \mathbf{z}_\tau}{d\tau} &= v_\theta(\mathbf{z}_\tau, \tau), \quad \mathbf{z}_\tau = (1-\tau)\cdot\mathbf{z}_0 + \tau\cdot\mathbf{z}_1, \quad \tau \in [0,1] \\
\mathcal{L}_{\mathrm{FM}} &= \mathbb{E}_{\tau, \mathbf{z}_0, \mathbf{z}_1} \big\| v_\theta(\mathbf{z}_\tau, \tau, \mathbf{C^I}, \mathbf{C^I_T}) - (\mathbf{z}_1 - \mathbf{z}_0) \big\|^2,
\end{align}
where $\theta$ denotes model parameters, $\tau$ is the flow step, and $\mathcal{L}_{\mathrm{FM}}$ is the flow-matching training loss.
The two duration predictors are trained together with the backbone using the following losses:
\begin{equation}
   \mathcal{L}_{\text{dur-clip}} = \mathbb{E}\|d_c - \hat{d}_g\|^2, \quad
   \mathcal{L}_{\text{dur-seq}} = \mathbb{E}_i\|(d_s)_i - (\hat{d}_s)_i\|^2,
\end{equation}
where $\hat{d}_g$ and $\hat{d}_s$ are ground-truth clip duration and sequence duration.
For NTA tasks, $\mathcal{L}_{\text{dur-seq}}$ is omitted.
In practice, $d_s$ and $\hat{d}_s$ are converted to frame numbers in the logarithmic domain to calculate $\mathcal{L}_{\text{dur-seq}}$, following FastSpeech2.
The final training loss is $\mathcal{L} = \mathcal{L}_{\mathrm{FM}} + \mathcal{L}_{\text{dur-clip}} + \mathcal{L}_{\text{dur-seq}}$.
During inference, classifier-free guidance (CFG) is employed to balance the trade-off between generated sample diversity and their fidelity to the input content: 
$
v_\theta^{\text{CFG}}(\mathbf{z}_\tau, \mathbf{C}^I, \mathbf{C}^I_T) 
= v_\theta(\mathbf{z}_\tau, \varnothing, \varnothing) 
+ w \cdot \Big( v_\theta(\mathbf{z}_\tau, \mathbf{C}^I, \mathbf{C}^I_T) 
- v_\theta(\mathbf{z}_\tau, \varnothing, \varnothing) \Big),
$
where $w$ is the guidance scale.


\section{Experimental Setup}

\paragraph*{Tasks and Data}
UniFlow-Audio is trained and evaluated on a series of public datasets.
Seven tasks are involved: TTS, SVS, T2A, T2M, SE, audio Super Resolution (SR) and V2A.
Among them, T2A and T2M are NTA tasks, while the rest are TA tasks.
Details of all training and evaluation data are demonstrated in \Cref{tab:data}.
A total of 7.7K hours of data are used for training, which is substantially less than that employed in UniAudio and AudioX.

\paragraph*{Task-Balanced Sampling}
As \Cref{tab:data} shows, different tasks' dataset sizes vary substantially due to discrepancies in collection difficulty and availability.
To prevent overexposure to small-scale datasets caused by random sampling, a straightforward approach is to adopt a task-based round-robin sampling strategy: sample data from each task in turn.
However, since the number of different task types is imbalanced (five TA tasks and two NTA tasks), task-based round-robin sampling disproportionately favors TA tasks during training, which may in turn affect the model's overall performance.
To this end, we upsample data from NTA tasks: T2M by 3 times and T2A by 2 times.
We refer to this sampling strategy as \emph{task-balanced sampling}.

\paragraph*{Training}
UniFlow-Audio is trained on eight A100 GPUs with a batch size on each GPU of 24.
We train three versions with different sizes: small, medium, and large.
Configuration and training details are in \Cref{subsec:backbone_details} and \Cref{subsec:training_inference_setup}.
The small version takes about 7 days to train, while the large version takes about 12 days.

\paragraph*{Evaluation Metrics}
For all tasks, both objective and subjective evaluation are conducted.
Since UniFlow-Audio is evaluated on a variety of tasks and datasets, we adopt task-specific commonly-adopted metrics, as illustrated in \Cref{sec:eval_metrics}.

\section{Results}

In this section, we first compare the performance of UniFlow-Audio with baselines on all tasks to evaluate the overall generation quality.
Then, we explore the effect of CFG scale on different tasks. 
Finally, we conduct ablation studies on our training and architecture design.

\subsection{Unified Audio Generation}

\begin{table}[tp]
    \setlength{\tabcolsep}{4pt}
    \centering
    \small
    \caption{Performance evaluation of UniFlow-Audio and baselines across all tasks.}
    \resizebox{\linewidth}{!}{
    \begin{tabular}{cccccc}
    \toprule
    \multirow{2}{*}{\textbf{Task}} & \multirow{2}{*}{\textbf{Model}} & \multicolumn{2}{c}{\textbf{Objective Evaluation}} & \multicolumn{2}{c}{\textbf{Subjective Evaluation}}\\
     & & \textbf{Metrics} & \textbf{Results} & \textbf{Metrics} & \textbf{Results}  \\
    \midrule
    \multirow{2}{*}{TTS} & NaturalSpeech 2\footnotemark~\citep{shen2024naturalspeech} & \multirow{2}{*}{WER$\downarrow$ $\mid$ SIM$\uparrow$} & 9.94 $\mid$ 34.8 & \multirow{2}{*}{MOS $\uparrow$ $\mid$ SMOS$\uparrow$} & 2.72 $\mid$ \textbf{3.43} \\
    & UniFlow-Audio &  & \textbf{3.09} $\mid$ \textbf{55.8} &  & \textbf{3.79} $\mid$ 3.21 \\
    \midrule
    \multirow{2}{*}{SVS} & DiffSinger~\citep{liu2022diffsinger} & \multirow{2}{*}{F0$\downarrow$ $\mid$ SA$\uparrow$} & \textbf{0.144} $\mid$ 58.0 & \multirow{2}{*}{MOS$\uparrow$ $\mid$ SMOS$\uparrow$} & \textbf{4.26} $\mid$ \textbf{4.43} \\
    & UniFlow-Audio &  & 0.147 $\mid$ \textbf{59.9} &  & 4.05 $\mid$ 4.31 \\
    \midrule
    \multirow{2}{*}{T2A} & AudioLDM 2~\citep{liu2024audioldm2} & \multirow{2}{*}{FD$\downarrow$ $\mid$ CLAP$\uparrow$} & 21.8 $\mid$ \textbf{0.476} & \multirow{2}{*}{OVL$\uparrow$ $\mid$ REL$\uparrow$} &  \textbf{3.57} $\mid$ 3.48 \\
    & UniFlow-Audio &  & \textbf{17.2} $\mid$ \textbf{0.476} &  & 3.41 $\mid$ \textbf{3.54} \\
    \midrule
    \multirow{2}{*}{T2M} & MusicGen~\citep{copet2023simple} & \multirow{2}{*}{FD$\downarrow$ $\mid$ CLAP$\uparrow$} & 29.5 $\mid$ \textbf{0.245} & \multirow{2}{*}{OVL$\uparrow$ $\mid$ REL$\uparrow$} & \textbf{3.45} $\mid$ 3.08 \\
    & UniFlow-Audio &  & \textbf{27.1} $\mid$ 0.241 &  & 3.37 $\mid$ \textbf{3.09} \\
    \midrule
    \multirow{2}{*}{SE} & DOSE~\citep{tai2023dose} & \multirow{2}{*}{PESQ$\uparrow$ $\mid$ STOI$\uparrow$} & 2.50 $\mid$ 0.931 & \multirow{2}{*}{MOS$\uparrow$} & 3.43 \\
    & UniFlow-Audio &  & \textbf{2.91} $\mid$ \textbf{0.944} &  & \textbf{4.76} \\
    \midrule
    \multirow{2}{*}{SR} & AudioSR~\citep{liu2024audiosr} & \multirow{2}{*}{LSD$\downarrow$} & 1.75 & \multirow{2}{*}{ MOS$\uparrow$} &  3.58 \\
    & UniFlow-Audio &  & \textbf{1.49} &  & \textbf{4.19} \\
    \midrule
    \multirow{2}{*}{V2A} & DiffFoley~\citep{luo2023diff} & \multirow{2}{*}{IB$\uparrow$ $\mid$ SYNC$\downarrow$} & 22.7 $\mid$ \textbf{922} & \multirow{2}{*}{ OVL$\uparrow$ $\mid$ SYNC$\uparrow$ } & 2.80 $\mid$ 2.94 \\
    & UniFlow-Audio &  & \textbf{28.6} $\mid$ 1145 &  & \textbf{3.61} $\mid$ \textbf{3.55} \\
    \bottomrule
    \end{tabular}
    }
    \label{tab:main_results}
\end{table}
\footnotetext{We use the open-source version \url{https://huggingface.co/amphion/naturalspeech2_libritts}.}


The comparison between UniFlow-Audio and prior works is demonstrated in \Cref{tab:main_results}.
For each task, we select a model from prior single-task works whose architecture and training data are closely aligned with our setting, while also demonstrating competitive performance.
Although unified audio generation has been explored (see \Cref{subsec:unified_audio_generation}), UniAudio~\citep{yang2024uniaudio} is trained on much larger data ,while AudioX~\citep{tian2025audiox} can only handle audio and music generation tasks.
Therefore, we do not incorporate them for comparison.
Except for LM-based MusicGen~\citep{copet2023simple}, all other baseline models adopt the diffusion architecture.
UniFlow-Audio achieves at least comparable performance to baselines and significantly outperforms baselines on TTS, SE and SR.
For TTS, both objective and subjective scores show the better synthesis quality of UniFlow-Audio.
Compared with NaturalSpeech 2, we observe greater diversity in the prosody of the synthesized speech, which leads to lower subjective scores for speaker similarity.
For SVS, a specified vocoder with high reconstruction quality is used in DiffSinger, while UniFlow-Audio uses a universal VAE, resulting in slightly lower singing synthesis quality on soprano samples.
For other tasks, UniFlow-Audio performs quite competitively with training only on limited public datasets.
In comparison, MusicGen~\citep{copet2023simple} was trained on private datasets, and DiffFoley is trained on VGGSound~\citep{chen2020vggsound}, which is ten times the size of VisualSound.

\begin{table}[ht]
    \centering
    \small
    \caption{Generation performance across different model sizes.}
    \resizebox{\linewidth}{!}{
    \begin{tabular}{c|c|ccccccc}
    \toprule
    \multirow{2}{*}{\textbf{Model}} & \multirow{2}{*}{\textbf{\shortstack{\# Trainable\\ Params}}} & TTS & SVS  & T2A & T2M & SE & SR & V2A \\
     &  & WER$\downarrow$ & SA$\uparrow$ & FD$\downarrow$ & FD$\downarrow$ & PESQ$\uparrow$ & LSD$\downarrow$ & IB$\uparrow$ \\
    \midrule
    Prior Works & - & 9.94 & 58.0 & 21.8 & 29.5 & 2.50 & 1.75 & 22.7 \\ 
    \midrule
    UniFlow-Audio small & 208M & 3.23 & 56.6 & 19.7 & \textbf{26.2} & 2.60 & 1.58 & 25.5 \\
    UniFlow-Audio medium & 395M & \textbf{3.03} & 58.4 & 17.8 & 26.6 & 2.72 & 1.53 & 26.5 \\
    UniFlow-Audio large & 847M & 3.09 & \textbf{59.9} & \textbf{17.2} & 27.1 & \textbf{2.91} & \textbf{1.49} & \textbf{28.6} \\
    \bottomrule
    \end{tabular}
    }
    \label{tab:model_size_result}
\end{table}

We also explore the effect of model size on the generation performance.
\Cref{tab:model_size_result} shows that UniFlow-Audio achieves competitive performance even with relatively few parameters.
UniFlow-Audio small, with only 208M trainable parameters, already outperforms baseline models across most tasks.
This demonstrates that UniFlow-Audio is parameter-efficient, delivering strong results without relying on excessively large model sizes.
We assume it can be attributed to the benefit of multi-task training since there is intrinsic commonality in the knowledge required by different tasks.
For example, TTS and SVS both require generating vocal from phoneme inputs, while T2A and T2M both require generating sound from coarse textual descriptions.
In contrast, other universal generation models, \textit{i.e.}, UniAudio~\citep{yang2024uniaudio} and AudioX~\citep{tian2025audiox}, both contain more than 1B parameters.
Although medium and large model versions further improve performance on certain tasks, the performance gap between the small model and its larger counterparts remains moderate.

\subsection{Effect of CFG and Inference Steps}
\begin{figure}[ht]
    \centering
    \includegraphics[width=0.95\linewidth]{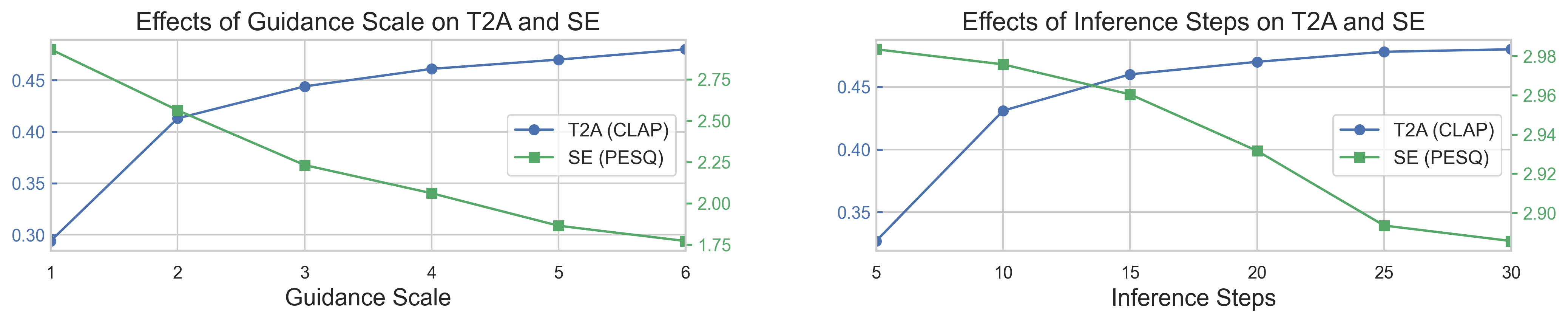}
    \caption{The effect of guidance scale (left) and inference steps (right) on generation performance of typical tasks. When analyzing one factor, the other is kept fixed.}
    \label{fig:cfg_steps}
\end{figure}
\label{subsec:cfg_step_effect}

We further investigate the impact of two key hyper-parameters in flow matching on generation performance: the guidance scale and the number of inference steps.
Interestingly, we observe two distinct patterns across all tasks: SE and SR fall into one pattern, while the remaining tasks follow another.
We take SE and T2A as representative tasks of the two patterns and report their CLAP and PESQ scores, with higher values indicating better performance for both metrics.
Results are presented in \Cref{fig:cfg_steps}.

For the T2A task, the effects of the guidance scale and inference steps are consistent with typical findings in diffusion-based models: larger guidance scales and more inference steps yield steady performance improvements.
This is expected, as stronger guidance provides more effective conditioning from the textual description, while more inference steps allow smaller step sizes in the denoising trajectory, which improves fidelity by reducing error accumulation.
However, SE exhibits a sharp performance decline as the guidance scale increases, with PESQ dropping from 2.9 to 1.75.
We attribute this to the characteristics of SE: the input inherently contains both signal and noise.
Stronger guidance thus amplifies not only the signal but also the noise, leading to reduced perceptual quality in the generated speech.
In contrast, the input of T2A is a textual description without ``noise'', so all information should ideally be reflected in the generated audio.
Regarding inference steps, increasing the number of steps is also detrimental to the performance, although the effect is considerably smaller than that of the guidance scale ($2.98 \rightarrow 2.89$).
This degradation may also stem from the fact that SE inputs contain both signal and noise.
With more inference steps, residual noise can accumulate through the iterative denoising process, slightly reducing the perceptual quality.

\subsection{Ablation Studies}

In this section, we conduct ablation studies to validate several components of UniFlow-Audio: 1) architecture design, including dual-fusion and layerwise fusion mechanisms, and 2) the task-balanced data sampling strategy.

\begin{table}[htp]
    \centering
    \small
    \caption{Ablation results on the architecture design and data sampling strategies of UniAudio-Flow. The best results are highlighted in bold, while the second-best are underlined.}
    \begin{tabular}{l|ccccc|cc}
    \toprule
    \multirow{3}{*}{\textbf{Setting}} & \multicolumn{5}{c|}{Time Aligned} & \multicolumn{2}{c}{Non Time Aligned} \\
    \cmidrule{2-8}
     & TTS & SVS  & SE & SR & V2A & T2A & T2M \\
     & WER$\downarrow$ & SA$\uparrow$ & PESQ$\uparrow$ & LSD$\downarrow$ & IB$\uparrow$ & FD$\downarrow$ & FD$\downarrow$\\
    \midrule
    UniFlow-Audio-small & \textbf{3.23} & \underline{56.6} & \underline{2.60} & \underline{1.58} & \underline{25.5} & \textbf{19.7} & \textbf{26.2} \\
    \midrule
    \hspace{1em} w. cross attention & 27.6 & 55.0 & 1.10 & 2.42 & 24.5 & 30.1 & 37.2 \\
    \hspace{1em} w. double fusion & 3.42 & \textbf{56.9} & \textbf{2.65} & \underline{1.58} & \underline{25.5} & 22.3 & 30.5\\
    \midrule
    \hspace{1em} w. input fusion & 42.0 & 41.8 & 1.07 & 1.59 & 13.7 & \underline{20.9} & 28.7 \\
    \midrule
    \hspace{1em} w/o. balanced sampling & \underline{3.30} & 56.5 & 2.54 & \textbf{1.53} & \textbf{26.0} & 22.9 & \underline{27.9} \\
    \bottomrule
    \end{tabular}
    \label{tab:ablation}
\end{table}

\begin{figure}
    \centering
    \includegraphics[width=0.95\linewidth]{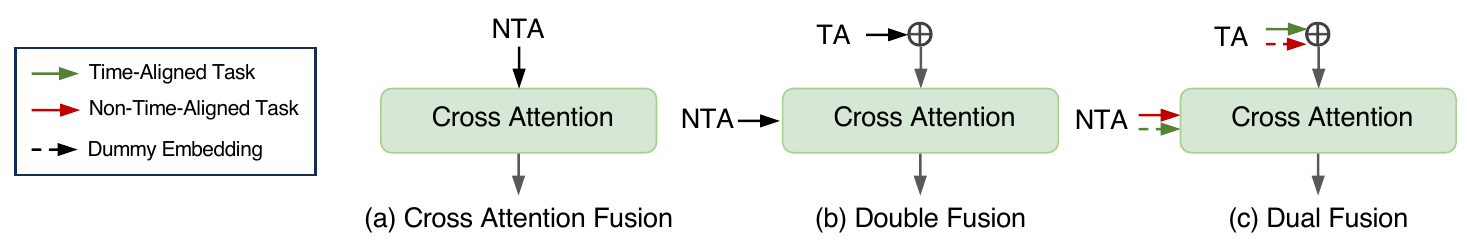}
    \caption{Illustration of different fusion mechanisms, best viewed in color. In the dual fusion subfigure, green and red indicate the flow in TA and NTA tasks, respectively, while the dashed line represents dummy embeddings. For instance, in TA tasks, the NTA content embeddings are replaced with dummy ones.}
    \label{fig:dual_fusion}
\end{figure}
\subsubsection{Benefits of Dual-Fusion Transformer}


To validate the effectiveness of our proposed dual-fusion mechanism, we replace it with alternative fusion strategies and compare their generation performance.
As \Cref{fig:dual_fusion} illustrates, we investigate two alternative fusion mechanisms: \emph{cross-attention fusion} and \emph{double fusion}.
Cross-attention fusion is the most straightforward approach, where all contents are fused with the audio latent via cross-attention, similar to AudioLDM2~\citep{liu2024audioldm2}.
Double fusion resembles our proposed dual fusion mechanism but differs in one aspect: content embeddings both before and after duration adaptation are fed into the backbone, regardless of the task type.
In contrast, in dual fusion, ineffective content embeddings based on task types are set to dummy embeddings.
This design may introduce interference between the learning of different task types.
In contrast, the dual fusion mechanism employs dummy embeddings, which provide better guidance for the model to attend to different sources depending on the task type, thereby mitigating such interference.

The upper half of \Cref{tab:ablation} reports the results of alternative content fusion mechanisms, which are consistent with our assumptions.
Although cross-attention has shown strong performance in prior T2A and T2M studies~\citep{liu2024audioldm2}, applying it directly to a mixture of task types results in poor performance.
Even on non-time-aligned T2A and T2M tasks, its performance is significantly worse than that of dual fusion, suggesting that the presence of rich time-aligned data adversely affects models based on cross-attention.
Compared with double fusion, dual fusion achieves similar performance on time-aligned tasks, while substantially outperforming it on non-time-aligned tasks.
This demonstrates the effectiveness of the dummy embedding design.
As described in \Cref{subsec:dual_fusion}, for non-time-aligned tasks, the duration used for content expansion is a dummy value.
Consequently, the incorporation of expanded content embeddings into the generation process acts as noise.

\subsubsection{Benefits of Block-Wise Fusion}

To further validate the architectural design, we examine the effect of fusing time-aligned content embeddings only at the input layer, referred to as \emph{input fusion}.
This follows the design of F5-TTS~\citep{chen2025f5} and FlowSep~\citep{yuan2025flowsep}.
As shown in the middle row of \Cref{tab:ablation}, input fusion leads to a substantial performance drop on time-aligned tasks.
Since content embeddings are integrated via cross-attention in each DiT block, injecting time-aligned inputs solely at the input layer makes their influence much weaker than that of non-time-aligned inputs.
Consequently, non-time-aligned tasks are only marginally affected, while the performance on time-aligned tasks degrades significantly.
In contrast, UniFlow-Audio employs \emph{block-wise fusion}, where time-aligned content embeddings are injected into each DiT block.
This progressive fusion allows richer interactions between time-aligned content and audio latents, and proves essential for achieving robust performance across different task types.

\subsubsection{Benefits of Task-Balanced Sampling}

Finally, we investigate the impact of the proposed task-balanced data sampling strategy.
As shown in the last row of \Cref{tab:ablation}, removing balanced sampling (\emph{w/o balanced sampling}) results in degraded performance on non-time-aligned tasks (T2A and T2M), while performance on time-aligned tasks remain relatively stable.
This aligns with the number of datasets from different task types: under the original round-robin sampling strategy, time-aligned tasks are overrepresented.
Without explicit balancing, the model is more exposed to time-aligned tasks, which amplifies the influence of time-aligned content input.
In contrast, the task-balanced sampling strategy ensures that each task type is adequately represented, mitigating the effects of task imbalance and leading to more consistent and reliable performance across both time-aligned and non-time-aligned tasks.

\section{Limitations}
Despite unifying TA and NTA audio generation within a flow-matching-based NAR framework, UniFlow-Audio has several limitations.
First, tasks involving multiple TA/NTA inputs, such as voice conversion (source speech + target speaker utterance), are not explored.
Second, the model's generalization to unseen tasks or input modalities, similar to the zero-shot generalization capabilities of LLMs, has not been investigated.
Third, the data and model size have not been scaled.
Except for T2M, most tasks have under 1,000 hours of training data.
Finally, UniFlow-Audio currently focuses on single-stream audio generation, while multi-stream or multi-source generation (e.g., TTS with background music) remains largely underexplored.

\section{Conclusion}
\label{sec:conclusion}
We present UniFlow-Audio, a flow-matching-based universal audio generation framework that unifies both TA and NTA tasks within a single NAR model.
By introducing a dual-fusion mechanism with block-wise integration, UniFlow-Audio effectively combines TA and NTA features without cross-task interference.
The model leverages shared knowledge across multiple modalities, including text, audio, and vision, to enhance generation performance through unified audio modeling.
Extensive experiments demonstrate that, even with limited training data and moderate model size (as small as 200M trainable parameters), UniFlow-Audio achieves competitive performance across diverse tasks, highlighting its potential as a foundation model for unified NAR audio generation.

\newpage

\section*{Ethics and Reproducibility Statement}
The authors have read and adhere to the ICLR Code of Ethics.
This work does not involve human subjects, identifiable private data, or harmful applications.
All datasets used are publicly available and were used in accordance with their original licenses and intended purposes.
No external sponsorship or conflict of interest influenced the design or conclusions of this work.

All code and source files are provided in the supplementary material and will be publicly released.

\bibliography{iclr2026_conference}
\bibliographystyle{iclr2026_conference}

\appendix

\section{Data Details}

\begin{table}[htpb]
    \centering
    \small
    \caption{Training and evaluation data details of UniFlow-Audio.}
    \begin{tabular}{c|ll|c}
    \toprule
    \textbf{Task} & \textbf{Training} & \textbf{Evaluation} & \textbf{Training Duration / h} \\
    \midrule
    TTS & \multicolumn{2}{c|}{LibriTTS~\citep{zen2019libritts}} & 555 \\ 
    \midrule
    SVS & \multicolumn{2}{c|}{M4Singer~\citep{zhang2022m4singer}} & 30 \\ 
    \midrule
    T2A & \multicolumn{2}{c|}{AudioCaps~\citep{kim2019audiocaps}} & 253 \\ 
    \midrule
    \multirow{4}{*}{SE} & LibriTTS+Wham!& \multirow{4}{*}{\shortstack{VoiceBank+Demand \\ \citep{botinhao2016investigating}}}& 460\\
    & VCTK+Wham!& & 44\\
    & LJSpeech+Musan& & 24\\
    & VoiceBank+Demand& & 10\\
    \midrule
    \multirow{4}{*}{SR} & HQ-TTS & VCTK & 85\\
     & MUSDB & \multirow{2}{*}{MUSDB}& 47\\
     & MoisesDB &  & 26\\
     & FreeSound & ESC & 158\\
    \midrule
    T2M & MSD~\citep{mcfee2012million} & MusicCaps~\citep{agostinelli2023musiclm} & 5789 \\
    \midrule
    V2A & \multicolumn{2}{c|}{VisualSound~\citep{viertola2025temporally}} & 236 \\
    \midrule
    Total & \multicolumn{2}{c|}{-} & 7717 \\
    \bottomrule
    \end{tabular}
    \label{tab:data}
\end{table}

UniFlow-Audio is trained and evaluated on a series of public datasets.
Details of all training and evaluation data are shown in \Cref{tab:data}.
For TTS and SVS, we use the official training / validation / test splits of LibriTTS and M4Singer.
Details of other datasets are described in the following:
\paragraph*{T2A} 
The official training subset of AudioCaps is used for T2A training.
Each sample contains 5 captions in the test subset.
Following TANGO~\citep{ghosal2023text}, we randomly select one caption per sample for evaluation, and we use the same selected captions as in their setup.
\paragraph*{T2M} For T2M, we use songs from MSD~\citep{mcfee2012million} combined with LP-MusicCaps-MSD~\citep{doh2023lp} captions as the training data. 
The original song in MSD can be as long as 14 minutes.
During training, we randomly crop 10 seconds for training.
The widely-used benchmark MusicGen~\citep{copet2023simple} is used for evaluation.
\paragraph*{SE} For SE, we utilize the method in URGENT challenge~\citep{zhang2024urgent}
to simulate noisy speech.
The clean speech datasets include LibriTTS, VCTK Corpus~\citep{yamagishi2019cstr} and LJSpeech~\citep{ljspeech17}, while the noise datasets contain WHAM!~\citep{wichern2019wham} and noise subset of Musan~\citep{snyder2015musan}.
Room Impulse Rresponses (RIRs) dataset for simulation is the RIRs dataset in \cite{ko2017study}.
We choose VoiceBank+Demand~\citep{botinhao2016investigating} for both train and evaluation, which is widely used as a benchmark in SE. 
\paragraph*{SR}
For SR, we mainly follow the setup of AudioSR~\citep{liu2024audiosr}, while prioritizing the available sources for ease of collection.
The training datasets include MUSDB~\citep{rafii2019musdb18}, MoisesDB~\citep{pereira2023moisesdb}, HQ-TTS~\citep{liu2022voicefixer} and FreeSound~\citep{mei2024wavcaps}, while the evaluation uses ESC-50~\citep{piczak2015esc}, VCTK-test \citep{liu2022neural}, and MUSDB.
All high-quality recordings are first resampled to 24~kHz.
Since our VAE is designed to process 24~kHz audio, we choose a cutoff range of [2,6]~KHz for the downsampled audio. Based on the method introduced in NVSR~\citep{liu2022neural}, we then apply the low-pass filter within this range to simulate low-high resolution audio pairs.
\paragraph*{V2A} For V2A, since the widely used VGGSound~\citep{chen2020vggsound} dataset is constructed from in-the-wild videos without ensuring high audio-video correspondence, it includes a considerable amount of modality-mismatched samples where the video and audio are not semantically related.
This limitation is detrimental to training stability and the inherent irrelevance is harmful to the performance.
Therefore, we adopt the smaller but better audio-visual aligned VisualSound~\citep{viertola2025temporally} for both training and evaluation, which is curated based on ImageBind scores~\citep{girdhar2023imagebind} to identify videos with poor audio-visual correspondence.

\section{Evaluation Metrics}
\label{sec:eval_metrics}
\paragraph*{TTS} Following~\citep{speechx}, we use Word Error Rate (WER)\footnote{\url{https://huggingface.co/nvidia/stt_en_conformer_transducer_xlarge}} as an objective metric to evaluate the accuracy of generated speech with respect to the given transcription, and Speaker Similarity (SIM)\footnote{\url{https://huggingface.co/nvidia/speakerverification_en_titanet_large}} to assess the consistency of speaker characteristics between the generated and reference speech. For subjective evaluation, we employ the Mean Opinion Score (MOS) to measure overall speech naturalness and the Similarity MOS (SMOS) to assess perceived speaker similarity.
\paragraph*{SVS} Following \cite{wu2024toksing}, we use root mean square error of fundamental frequency (F0) and semitone accuracy (SA)\footnote{\url{https://github.com/espnet/espnet/blob/master/egs2/TEMPLATE/svs1/svs.sh\#L1171}} for objective evaluation.
Same as TTS, MOS and SMOS are used as subjective metrics for accessing singing quality and singer similarity.
\paragraph*{T2A \& T2M} Following previous T2A and T2M studies~\citep{liu2024audioldm2}, we adopt Frechet Distance (FD) and CLAP score for audio and music generation evaluation.
FD measures the similarity of the distribution between generated and reference audio based on PANNs CNN14~\citep{kong2020panns} features, while CLAP score serves as a reference-free metric that captures the semantic alignment between textual descriptions and generated audio. 
\paragraph*{SE} Following ~\cite{tai2023dose}, we choose Perceptual Evaluation of Speech Quality (PESQ) and Short-Time Objective Intelligibility (STOI) for SE evaluation.
PESQ measures perceptual speech quality, and STOI estimates speech intelligibility.
\paragraph*{SR} Following previous studies~\citep{liu2024audiosr,liu2022neural}, we adopt Log-Spectral Distance (LSD) for objective evaluation.
LSD measures the discrepancy between the original high-frequency audio and the generated audio.
Note that the baseline model AudioSR generates 48~kHz audio, while ours operates at 24~kHz.
For fair comparison, AudioSR outputs are downsampled to 24~kHz before evaluation.

\paragraph*{V2A} Following \cite{viertola2025temporally}, we evaluate V2A performance using ImageBind~\citep{girdhar2023imagebind} (IB) and Synchformer~\citep{iashin2024synchformer} (SYNC).
IB measures semantic modality consistency by computing the cosine similarity between audio and video embeddings.
SYNC assesses synchronization based on temporal offsets between audio and visual modality estimated by Synchformer.

\section{Task Instructions}
For each task, we prompt the LLM to generate $10$ instructions ranging from simple to complex. 
These instructions span from basic definitions of the task to detailed specifications of task requirements. 
Table~\ref{tab:task_instruction} presents $3$ examples of simple, medium, and complex instructions.

\label{sec:task_instructions}

\begin{table}[ht]
    \centering
    \small
    \caption{Examples of detailed task instructions.
    }
    \begin{tabular}{c|l}
    \toprule
    \midrule
    \multirow{5}{*}{\textbf{TTS}}
	&Produce human-like speech from phoneme inputs and speaker representations. \\
    \cline{2-2}
    &	Generate natural speech from speaker embeddings and phoneme sequences while maintaining  \\ 
    & accurate pronunciation. \\
    \cline{2-2}
	&Convert phoneme sequences into natural speech using speaker embeddings, with precise articulation \\ 
    & of words and adaptation to the textual emotional content. \\
	
    \midrule
    \midrule
    
    \multirow{6}{*}{\textbf{T2A}} & Generate an audio clip based on the given text description. \\
    \cline{2-2}
    & Synthesize an audio signal from the given text, ensuring the fidelity of sound event representation \\ &and the naturalness of the audio output. \\
    \cline{2-2}
    & Convert the given text into a natural-sounding audio clip, maintaining high fidelity in sound event  \\ 
    & reproduction (volume, positioning, timing, repetition) and ensuring realistic scene acoustics and  \\ 
    & event relationships.\\
    
    \midrule
    \midrule
    
    \multirow{6}{*}{\textbf{SVS}} &   Render a singing performance from musical notation, including phonemes, notes, durations, and slurs. \\
	\cline{2-2}
	&   Produce a singing voice rendering derived from the notated score that maintains parametric fidelity to \\ 
    &   the given phonemes, notes, durations, and slurs.  \\
	\cline{2-2}
	&   Synthesize a singing voice that matches the input musica score's specifications (phonemes, notes,  \\ 
    &   durations, slurs) while adapting phoneme durations for natural flow and preserving textual \\
    &    emotional tone.\\

    \midrule
    \midrule
    
\multirow{5}{*}{\textbf{SE}}&	Enhance noisy speech signals by reducing background noise and reverberation. \\ 
\cline{2-2}
	&Improve degraded speech quality by suppressing noise and reverberation while preserving natural  \\ 
    &voice characteristics. \\
    \cline{2-2}
	&Enhance speech signals by dynamically suppressing diverse noise types (environmental/mechanical) \\ 
    & and reverberation, preserving tonal qualities and timbre across varying SNR conditions. \\
    
    \midrule
    \midrule
    
\multirow{6}{*}{\textbf{SR}}&	Enhance audio quality by increasing its sampling rate or resolution. \\
	\cline{2-2}
	&Convert low-sampling-rate audio to high-resolution output, recovering lost high-frequency components \\ 
    & and subtle sonic characteristics. \\
    \cline{2-2}
	&Upsample low-resolution audio signals to higher sampling rates while preserving original signal details\\ 
    & and recovering high-frequency components without introducing audible artifacts.  \\
    
	\midrule
    \midrule
    
\multirow{5}{*}{\textbf{V2A}}&	Generate high-fidelity audio synchronized to video.  \\
	\cline{2-2}
	&Produce high-quality audio that matches the video’s scene, with accurate timing, spatial positioning,  \\ &and realistic sound properties. \\
    \cline{2-2}
	&Generate high-fidelity audio for the video, ensuring strict temporal alignment, correct spatial direction,  \\ 
    &loudness, and frequency of sounds, while maintaining realism and coherence with visual content. \\

	\midrule
    \midrule
    
\multirow{6}{*}{\textbf{T2M}}&	Develop a music clip that precisely matches the textual description in all aspects. \\ 
\cline{2-2}
	&Produce a musical piece that faithfully represents the given description, incorporating all \\ 
    & specified instruments, intended emotions, genre characteristics, and vocal properties. \\
\cline{2-2}
	&Generate a musical output that perfectly matches the provided text, incorporating the exact instruments \\ &mentioned, upholding authentic stylistic qualities, and delivering the desired emotional impact.   \\
    & If vocals are required, precisely implement the described gender, age, vocal properties, and singing manner. \\
    \bottomrule
    \end{tabular}
    \label{tab:task_instruction}

\end{table}

\section{Architecture \& Hyper-Parameters}
\label{sec:arch_training_details}

\subsection{Waveform-based VAE}
\label{subsec:vae_details}

\begin{table}[h]
    \centering
    \small
    \caption{Datasets used for training the waveform-based VAE.}
    \label{tab:vae_data}
    \begin{tabular}{ll}
        \toprule
        \textbf{Domain} & \textbf{Datasets} \\
        \midrule
        Speech  & AISHELL-3~\citep{shi2021aishell}, TTS-HQ, LJSpeech, LibriTTS, VCTK \\
        Singing & OpenSinger~\citep{huang2021multi}, M4Singer, OpenCpop~\citep{wang2022opencpop}, PopCS~\citep{liu2022diffsinger} \\
        Music   & MUSDB, MoisesDB, MusicCaps \\
        General Audio & AudioSet~\citep{gemmeke2017audio}\\
        \bottomrule
    \end{tabular}
\end{table}
The VAE adopts a fully-convolutional architecture with residual 1D blocks and Snake activations, following the design from \cite{evans2025stable}.
The encoder maps raw waveforms into a compact latent sequence at a downsampling ratio of 480 with 128 channels, while the decoder mirrors the encoder by progressively upsampling the latent sequence with transposed convolution to reconstruct the waveform.
To achieve high-fidelity audio generation across different audio types, we train the VAE using a diverse set of datasets from multiple categories, including speech, singing, music, and general audio, with details provided in Table~\ref{tab:vae_data}.
The model is trained for 1M steps on this extensive collection of approximately 6000 hours data, where each audio clip is randomly cropped to 1.5s segments during training.


To measure the reconstruction quality of VAE, we evaluate the mean squared error (MSE) and signal-to-noise ratio (SNR) on held-out test sets.
As shown in Table~\ref{tab:vae_results}, our VAE achieves consistently lower MSE and higher SNR than the one in EzAudio~\citep{hai2024ezaudio}, which was only trained on AudioSet~\citep{gemmeke2017audio}. 

\begin{table}[htbp]
    \centering
    \small
    \caption{Reconstruction performance of VAE.}
    \label{tab:vae_results}
    \begin{tabular}{l|cc|cc}
        \toprule
        \multirow{2}{*}{\textbf{Domain}} & \multicolumn{2}{c|}{\textbf{Speech}} & \multicolumn{2}{c}{\textbf{Music}} \\
        & \textbf{MSE $\downarrow$} & \textbf{SNR (dB) $\uparrow$} & \textbf{MSE $\downarrow$} & \textbf{SNR (dB) $\uparrow$} \\
        \midrule
        EzAudio VAE  & $4.43 \times 10^{-5}$ & 17.06 & $1.13 \times 10^{-4}$ & 18.09 \\
        Ours & $3.84 \times 10^{-5}$ & 17.63 & $8.42 \times 10^{-5}$ & 19.27 \\
        \bottomrule
    \end{tabular}
\end{table}

\subsection{Flow Matching Backbone}
\label{subsec:backbone_details}
The diffusion step $\tau$ is processed by a multi-layer perceptron (MLP) to produce AdaLN scale and shift parameters for each Transformer block, conditioning the self-attention and FFN layers:
\begin{equation}
 \gamma_{\mathrm{SA}}, \beta_{\mathrm{SA}}, \alpha_{\mathrm{SA}}, \gamma_{\mathrm{FFN}}, \beta_{\mathrm{FFN}}, \alpha_{\mathrm{FFN}} = \mathrm{MLP}(\tau)
\end{equation}
We apply $\tanh$ to the scaling parameter $\alpha$ in AdaLN~\citep{peebles2023scalable} to improve the numerical stability during training:
\begin{align} 
     \mathbf{A}_{\text{norm}}  &= \gamma \cdot \mathrm{Norm}(\mathbf{A}) + \beta\\
     \mathbf{A} &= \tanh(1 - \alpha) \odot \mathrm{F}(\mathbf{A}_{\text{norm}})+\mathbf{A}_{\text{norm}}
\end{align}
\par

To mitigate the potential negative influence from $\mathcal{L}_{\text{dur-clip}}$ and $\mathcal{L}_{\text{dur-seq}}$, we apply gradient scaling to the duration predictors.
Specifically, we scale the gradients from the duration losses by a factor $\lambda$ before backpropagation, thereby reducing their influence on the model.
\begin{align*}
\tilde{x}  = \lambda \cdot  x+(1-\lambda) \cdot \mathrm{sg}(x)
\end{align*}
where $\mathrm{sg}(\cdot)$ represents stop gradient operator and $\lambda$ is set to 0.1.

\Cref{tab:model_config} summarizes the architectural configurations of different UniFlow-Audio versions.
Notably, the small variant contains only approximately 200M trainable parameters, yet it achieves competitive performance as shown in \Cref{tab:model_size_result}.
\begin{table}[h!]
    \centering
    \caption{Model configurations.}
    \begin{tabular}{lcccc}
        \toprule
        Model Size  & Depth & Embed Size & Num Heads & \# Total / Trainable Params \\
        \midrule
        Small  &   12    &  512         &8    &593M / 208M           \\
        Medium &   16    &  768         &12    &780M / 395M           \\
        Large  &   24    &  1024        &  16 & 1.2B / 847M        \\
        \bottomrule
    \end{tabular}
    \label{tab:model_config}
\end{table}

\subsection{Training \& Inference Setup}
\label{subsec:training_inference_setup}
UniFlow-Audio is trained using AdamW optimizer~\citep{loshchilov2017decoupled} with a constant learning rate of 5e-5 with a warmup step of 10K steps and a total training step of 400K steps.
To mitigate the negative impact of excessively long audio content sequence on training efficiency, we take a maximum of 5 second audio segments randomly during training for SE and SR.
During inference, we take an inference step of 25 by default.
Sway sampling~\citep{chen2025f5} is adopted to improve the generation performance.
During training, both TA and NTA content embeddings are randomly masked with a ratio of 0.2 to train conditional and unconditional generation simultaneously.
During inference, a CFG scale of 5.0 is adopted for tasks except SE and SR while CFG is not applied for these two tasks, due to the influence of CFG on them (see \Cref{subsec:cfg_step_effect}).

\section{LLM Usage}
LLMs were used as assistive tools in this work.
Specifically, they were employed to help with limited code writing and debugging, as well as for polishing the language of the paper.
The LLMs involved include mainstream models such as GPT, Claude, and Gemini.
These model were used for grammar correction, sentence restructuring, and enhancing overall readability.
All technical content, experimental design, results, and conclusions were authored and verified solely by the human authors.
LLMs did not contribute to the generation of ideas, methods, or data analysis.

\section{Subjective Evaluation Details}

\begin{figure}[ht]
    \centering
    \begin{subfigure}[t]{0.48\textwidth}
        \centering
        \includegraphics[height=4.5cm]{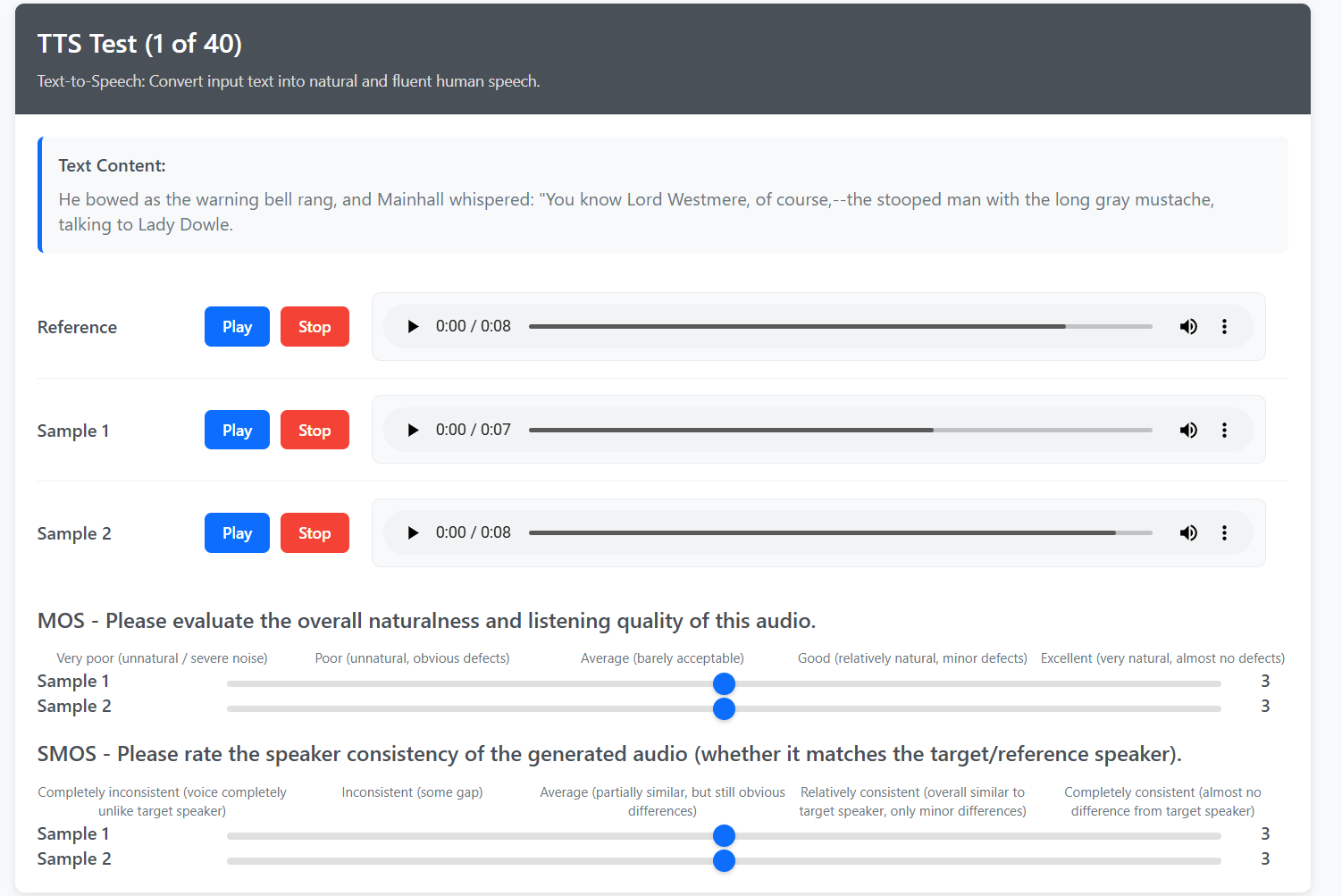}
        \caption{TTS \& SVS evaluation interface.}
        \label{fig:eval_tts}
    \end{subfigure}
    \hfill
    \begin{subfigure}[t]{0.48\textwidth}
        \centering
        \includegraphics[height=4.5cm]{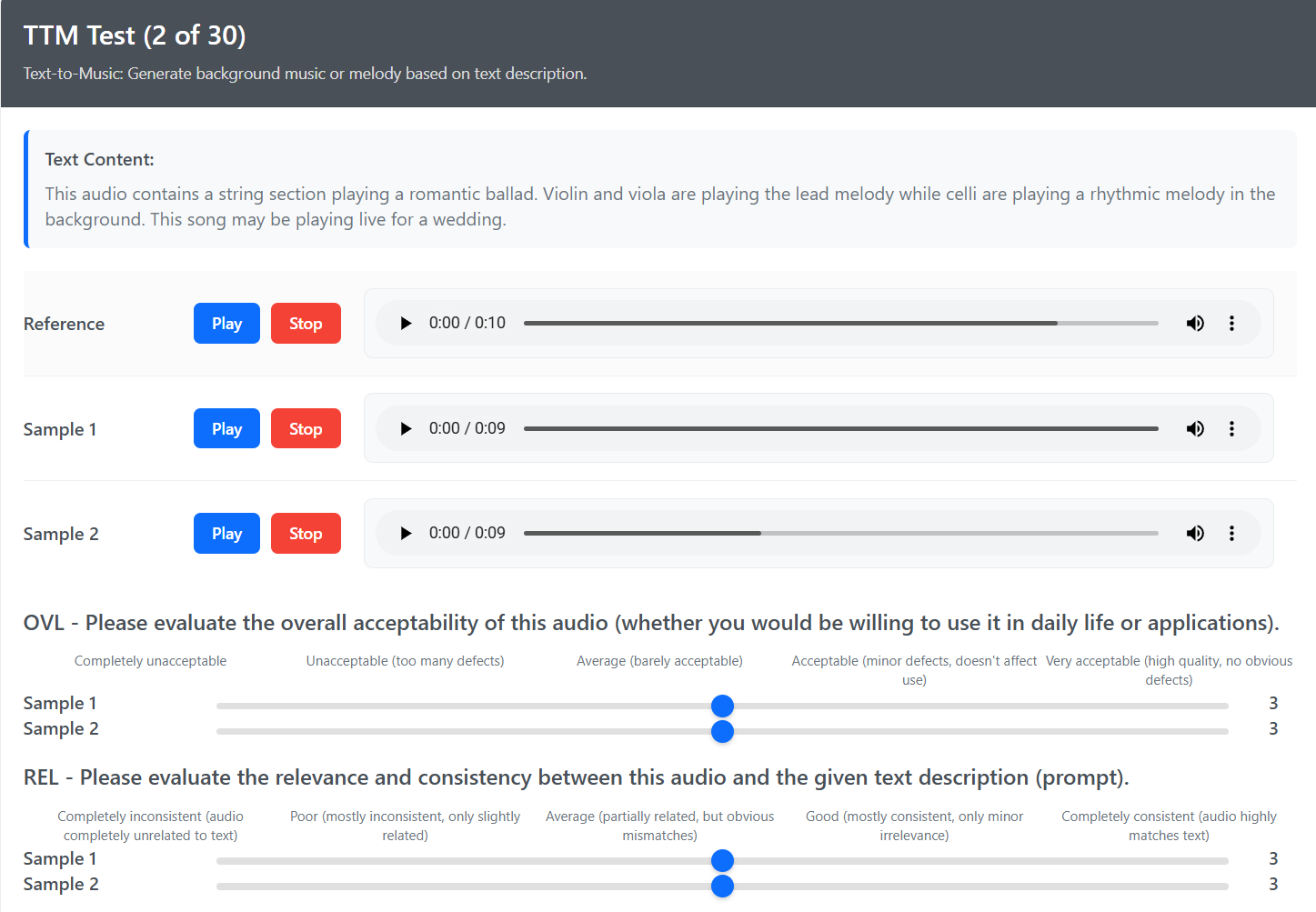}
        \caption{T2M \& T2A evaluation interface.}
        \label{fig:eval_ttm}
    \end{subfigure}
    
    \vspace{0.6em}
    
    \begin{subfigure}[t]{0.48\textwidth}
        \centering
        \includegraphics[height=4.5cm]{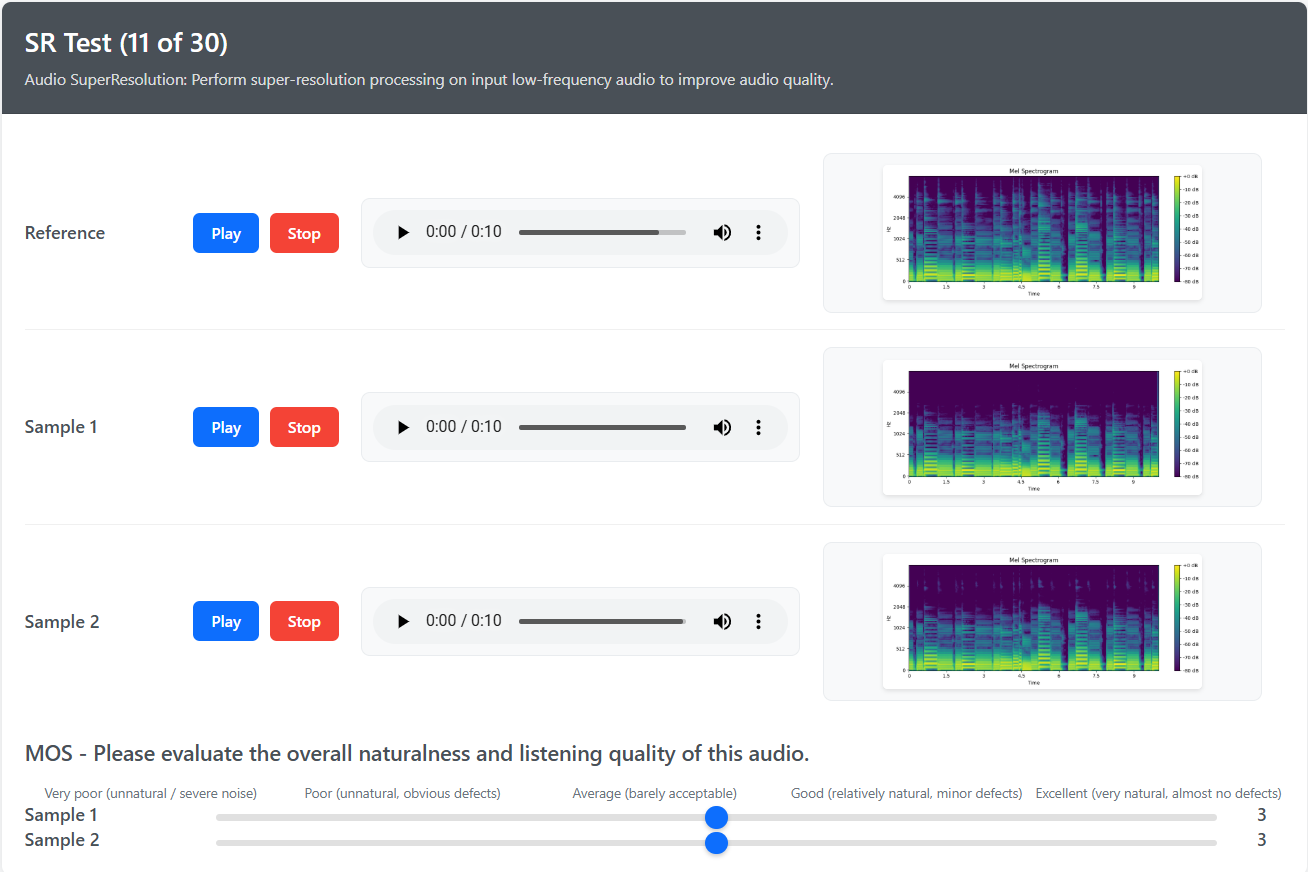}
        \caption{SR evaluation interface.}
        \label{fig:eval_sr}
    \end{subfigure}
    \hfill
    \begin{subfigure}[t]{0.5\textwidth}
        \centering
        \includegraphics[height=4.5cm]{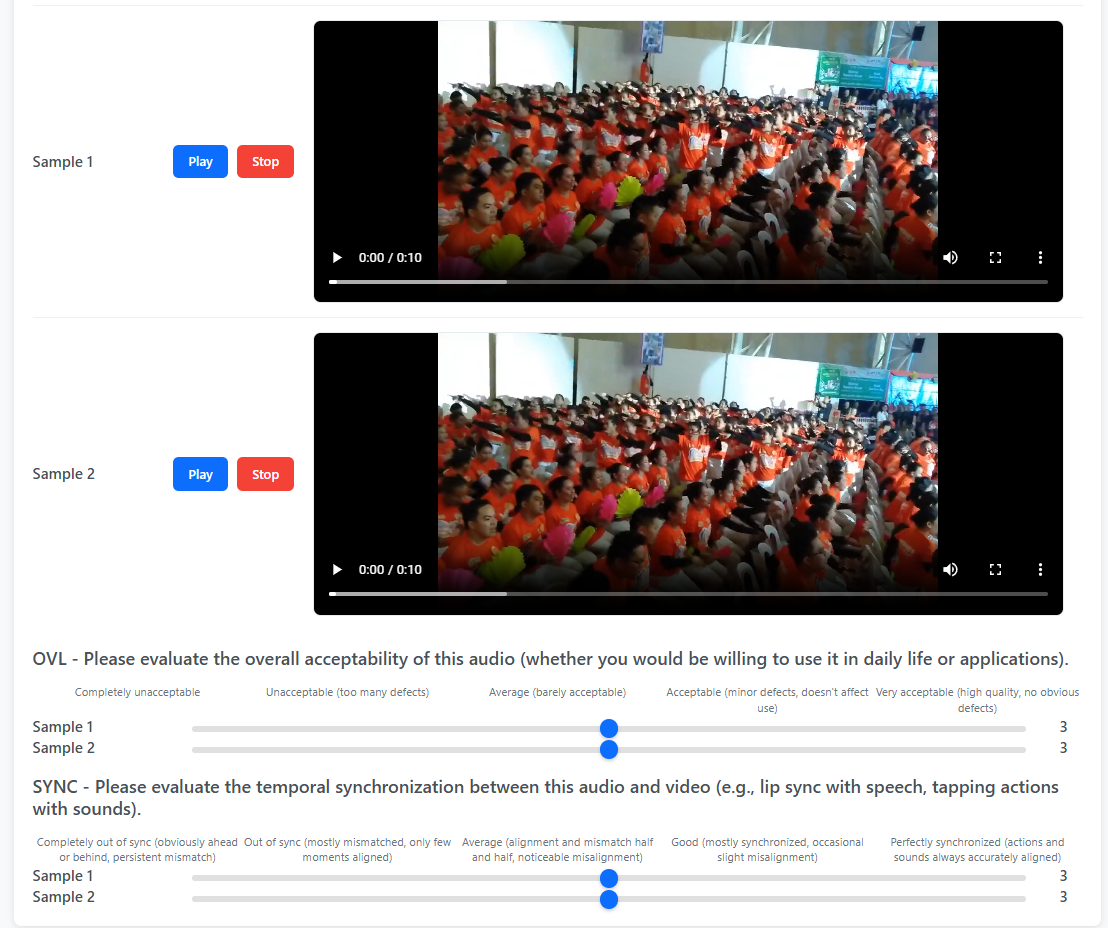}
        \caption{V2A evaluation interface.}
        \label{fig:eval_v2a}
    \end{subfigure}
    
    \caption{Screenshots of the subjective evaluation interfaces used in our experiments.}
    \label{fig:eval_interfaces}
\end{figure}


For all tasks, we conduct MOS-based subjective tests with explicit instructions for raters.
Each sample is rated on a 1--5 Likert scale.
We recruit ten raters with college-level education and normal hearing ability for subjective evaluation. 
Examples of the rating interface and detailed instructions are shown in \Cref{fig:eval_interfaces}.
Below we describe the setup for each task.

For \textbf{TTS} and \textbf{SVS}, we evaluate speech quality MOS (MOS) and speaker similarity MOS (SMOS).
For MOS, raters judge the overall naturalness and listening quality of the synthesized speech or singing voice.
For SMOS, raters judge whether the generated audio matches the target/reference speaker in terms of timbre-related characteristics, disregarding prosodic variations.

For \textbf{T2A} and \textbf{T2M}, we follow AudioGen~\citep{kreuk2022audiogen} and MusicGen~\citep{copet2023simple} to evaluate overall quality (OVL) and relevance (REL) to the input caption.

For \textbf{SE}, raters assess the intelligibility and naturalness
of enhanced speech.
Each output is presented together with its clean reference target, and the MOS scores reflect residual noise, processing artifacts, and overall listening quality.

For \textbf{SR}, the evaluation setup is identical to SE, except that each sample is additionally accompanied by a spectrogram visualization to facilitate judgments.

For \textbf{V2A}, we evaluate overall acceptability (OVL) and synchronization (SYNC) with the reference video.
In SYNC evaluation, the raters judge whether audio events are temporally aligned with visual cues such as lip movements, object impacts, or musical actions.

\end{document}

%% file: math_commands.tex
\usepackage{amsmath,amsfonts,bm}

\def\eqref#1{equation~\ref{#1}}

\def\1{\bm{1}}

\DeclareMathAlphabet{\mathsfit}{\encodingdefault}{\sfdefault}{m}{sl}
\SetMathAlphabet{\mathsfit}{bold}{\encodingdefault}{\sfdefault}{bx}{n}

